# Plasma waves excited at interface by femtosecond laser irradiation enabling formation of volume nanograting in glass

Yang Liao[1], Jielei Ni[1], Lingling Qiao[1], Min Huang[2], Yves Bellouard[3], Koji Sugioka[4], and Ya Cheng[1,*]

[1]*State Key Laboratory of High Field Laser Physics, Shanghai Institute of Optics and Fine Mechanics, Chinese Academy of Sciences, Shanghai 201800, China,*
[2]*State Key Laboratory of Optoelectronic Materials and Technologies, Sun Yat-sen University, Guangzhou 510275, China*
[3]*Mechanical Engineering Department, Eindhoven University of Technology, 5600 MB Eindhoven, The Netherlands*
[4]*RIKEN-SION Joint Research Unit, RIKEN - Advanced Science Institute, Center for Advanced Photonics, Hirosawa 2-1, Wako, Saitama 351-0198, Japan*

[*]Email: *ya.cheng@siom.ac.cn*

**Abstract**

Irradiation of intense ultrafast laser pulses in glasses can lead to formation of nanogratings whose periods are significantly smaller than the incident irradiation wavelength. The mechanism of the exotic phenomenon is still under debate. Here, we access the snapshots of morphologies in the laser affected regions in a porous glass which reveal the evolution of the formation of nanogratings with increasing number of laser pulses. Combined with further theoretical analyses, our observation provides important clues which suggest that excitation of standing plasma waves at the interfaces between areas modified and unmodified by the femtosecond laser irradiation plays a crucial role for promoting the growth of periodic nanogratings. The finding indicates that the formation of volume nanogratings induced by irradiation of femtosecond laser pulses is initiated with a mechanism similar to the formation of surface nanoripples.

Recently, the interaction of ultrafast laser pulses with condensed matter has attracted significant attention. It opens new horizons in strong field laser physics and attosecond science that has largely focused on gaseous media over the past two decades as well as new avenues for materials laser processing beyond the diffraction limit (i.e., with nanoscale spatial resolutions) and with three-dimensional (3D) capabilities [1-3]. Specifically, irradiation of intense ultrafast laser pulses inside dielectric materials has led to intriguing phenomena, such as formation of nanovoids and periodic nanogratings [4-7], glass constituent redistribution [8], nonreciprocal writing [9-11], as well as formation of new phases of matter [12]. Among them, the formation of nanogratings has been intensively investigated since its discovery, as the feature size of the nanograting is much less than the incident irradiation wavelength which provides promising potential for applications in both nanophotonics and nanofluidics [13,14]. At present, several models have been proposed to explain this phenomenon, including interference between the incident light field and the electric field of the bulk electron plasma wave excited by the incident light [5], the formation of randomly distributed nanoplasmas due to the initial inhomogeneous multiphoton absorption in the focal volume which self-organizes into periodic nanoplanes [6], and very recently, the coupling between attractive interaction and self-trapping of exciton-polaritons [15]. Nevertheless, a convincing explanation, that explains all reported experimental observations, has not yet been proposed. For direct observation of nanogratings formed in glasses by either scanning electron microscope (SEM) or atomic force microscope (AFM), the fabricated structures must be exposed to ambient air typically by cleavage or focused ion beam milling [16]. Afterwards, chemical etching is often employed to facilitate the observation of nanogratings. This post process, however, can modify or even destroy some fine features in the laser-modified zones (LMZs) [17]. We show that porous glass, which has been employed for fabricating 3D micro- and nanofluidic devices [18,19], provides us an ideal platform for investigating the evolution of nanogratings during its formation, because the irradiation by intense femtosecond laser of porous glass immersed in water can induce efficient ablation, resulting in formation of periodic nanocracks. As a result, the fine morphological features in the LMZ can directly be displayed under SEM after cleavage of the prepared samples. Therefore, the porous glass serves as an ideal three-dimensional (3D) recording medium that allows us to access the high-fidelity snapshots during the formation of nanogratings. Based on our observation, we suggest that excitation of plasma waves at the interface between the areas modified and unmodified by femtosecond laser irradiation plays a crucial role in initiating the formation of nanogratings.

The schematic of our experimental setup is illustrated in Fig. 1. In our experiment, high-silicate porous glass samples were used as the substrates, which were produced by removing the borate phase from phase-separated alkali-borosilicate glass in hot acid solution [20]. Details of the glass can be found elsewhere [14]. To induce the nanograting structures, a high-repetition regeneratively amplified Ti:sapphire laser (Coherent, Inc., RegA 9000) with a pulse duration of ~100 fs, a central wavelength of 800 nm and a repetition rate of 250 kHz was used. The Gaussian laser beam with an initial 8.8 mm diameter was passed through a ~3 mm-diameter aperture, i.e., only the central part of the laser beam was used because of its high homogeneity. A long-working-distance water-immersion objective (N.A. = 1.10) was employed for focusing the beam into the porous glass. The glass sample was fixed in a petri dish filled with distilled water. The femtosecond laser beam was focused in the volume of porous glass with two different focusing geometries. In the first geometry (Fig. 1(a)), the femtosecond laser beam was directly focused into the glass using the water-immersion objective; whereas in the second geometry (Fig. 1(b)), a narrow slit with a width of 0.6 mm and a length of 5 mm was placed above the objective lens, resulting in an expansion of the focal spot in the transverse direction. The nanogratings formed in the tracks written with different focusing geometries are schematically illustrated in the insets of Fig. 1. The purpose of the slit is to increase the focal spot size in the transverse direction and reduce the interaction length of the femtosecond laser with the glass in the longitudinal direction (i.e., along the Z axis as indicated in Fig. 1(a)) [21]. In all the experiments, the laser pulses were focused ~170 μm below the surfaces of glass samples. To characterize the morphologies of the embedded nanogratings, the fabricated samples were cleaved along the plane perpendicular to the writing direction to access the cross sections of the LMZs. The revealed nanograting structures were directly characterized using a SEM (Zeiss Auriga 40). Neither chemical etching nor annealing was used before the SEM examinations.

First, we present SEM images of the cross sections of nanostructures inscribed in the porous glass at a low pulse energy of 50 nJ and various scan velocities without the slit in Fig. 2. Under this laser intensity, an isolated subwavelength-width crack (SWC) with a transverse dimension of ~40 nm can be formed with sufficient number of laser pulses [14]. At the highest scan velocity of 25 mm/s, which corresponds to the least number of laser pulses deposited in the focal volume, a nanovoid of a diameter of ~300 nm was formed in the central area of focal spot (Fig. 2(a)). By reducing the scan velocity to 10 mm/s, it is found that some defects, which appear under the SEM to be hollow nanovoids with diameters larger than that of the original nanopores, started to gather along the optical axis of the incident laser in its backward direction, as indicated by the white arrows in Figs. 2(b)-2(c). Further reduction of the scan velocity leads to formation of a mature SWC as shown in Figs. 2(d)-2(e). The growth of the single SWC occurs in both backward and forward directions, as shown in Fig. 2(e).

When the pulse energy was increased to 120 nJ, which is well above the ablation threshold, we found that the induced phenomena are completely different from those in Fig. 2. At the beginning, a well defined LMZ can be formed around the focus, in which a large number of randomly distributed defects are formed (Fig. 3(a)). It should be stressed that at this stage, there is no any signature of nanograting formation in the LMZ. With the increasing number of pulses, it is clear that due to the modified optical property around the LMZ, the nonlinear propagation of the laser pulses leads to more complex behaviors such as single and multiple self-refocusing as shown in Figs. 3(b)-3(d), respectively. Interestingly, after a certain number of laser pulses, some signature of periodically arranged nanovoids array appears in Figs. 3(d) and 3(e). A close examination reveals that the periodic nanovoid chain is preferentially formed at the interface between the regions modified and unmodified by the femtosecond laser irradiation, as indicated by the black arrows in Figs. 3(d) and 3(e). Finally, after irradiated with more laser pulses, the nanovoids in Fig. 3(e) all grow into SWCs aligned in parallel with the laser propagation direction, leading to the formation of a periodic nano-crack array (PNCA) as shown in Fig. 3(f).

Since the morphological features in the LMZs can easily be distorted by nonlinear propagating effects such as self-focusing and defocusing, we designed another experiment to avoid this issue. It is now well known that by adding a narrow slit on the top of objective lens, the focal spot of laser beam can be transversely expanded in glass due to the diffraction effect of the slit [Fig. 1(b)] [21]. Since the laser intensity will have a more homogeneous distribution in the transverse direction, self-focusing can be effectively mitigated which results in only one well-defined focus in the longitudinal direction in the porous glass, as evidenced in Fig. 4. Here, the pulse energy was increased to 1.20 μJ as the slit used in front of the objective lens inevitably blocked a large part of the laser beam and caused significant loss. The results obtained at the scan velocities of 26 mm/s, 8 mm/s, 4.5 mm/s, 1.25 mm/s, which are presented in Figs. 4(a)-(d), respectively, show no signature of nanograting formation. However, in Fig. 4(d), a boundary between the regions affected and unaffected by the laser irradiation can be clearly seen. The following SEM image in Fig. 4(e) shows that at the interface, some periodically distributed nanovoids have been formed at a scan velocity of 0.5 mm/s. By further reducing the scan velocity to 0.35 mm/s, a few nanovoids develop into nanocracks whereas the remaining nanovoids are almost unchanged (Fig. 4(f)). At last, at a scan velocity of 0.01 mm/s, all the nanovoids develop into SWCs which leads to the formation of a perfect PNCA (Fig. 4(i)). These results establish an important link between the formation of volume nanograting in glass and surface nanoripples on various substrates observed previously, as the interface created in bulk glass can play the similar role for surface plasma wave excitation [22-23].

The above observation provides crucial guidance for theoretical modeling. First, we perform a finite difference time domain (FDTD) simulation for explaining the SWC formation as shown in Fig. 2. The nanovoid observed in Fig. 2(b) suggests that initially a spherical nanoplasma with a diameter of a few tens of nanometers has been produced by multiphoton ionization in the glass. Thus, we assume that the nanoplasma has a diameter of 40 nm and a plasma density of $N_e = 5 \times 10^{20}$ cm$^{-3}$. Such plasma density is chosen to best reproduce the experimental observations because direct measurement of the plasma density in the volume of glass is difficult. Since the assumed size of nanoplasma is much smaller than the laser wavelength, the laser field in our simulation is assumed to be a linearly polarized plane wave with a wavelength centered at 800 nm and a spectral width (FWHM) of ~30 nm. The relative dielectric constant of SiO$_2$ is assumed to be 2.1025. Using Drude model, the relative dielectric constant in the laser field including the effect of $N_e$ can be derived as below [23, 24]

$$\varepsilon = \varepsilon_m - \frac{N_e e^2}{\varepsilon_0 m^* m} \frac{1}{(\omega^2 + i\omega/\tau)}, \qquad (1)$$

where $\omega$ is the incident light frequency in vacuum, $\tau$ is the Drude damping time of free electrons, and $\varepsilon_0$, $m$, $m^*$ and $e$ are the relative dielectric constant in vacuum, the electron mass, optical effective mass of carriers and electron charge, respectively. In reality, the Drude damping time is varying as a function of free electron density and temperature [25]. For simplicity, we chose a fixed Drude damping time of 10 fs in our simulation as Ref. [26].

Figure 5(a) shows the simulated distribution of laser intensity around the spherical nanoplasma in the XZ plane with $Ne = 5\times10^{20}$ cm$^{-3}$, $\varepsilon = 0.81945+0.03712i$. From Fig. 5(a), we can see a clear enhancement of the laser intensity inside the nanoplasma as well as at its equator perpendicular to the incident electric-field vector. As explained in Ref. [27], the laser field inside the nanoplasma can be stronger than the incident field in an underdense nanoplasma ($0 < \text{Re}(\varepsilon) < \varepsilon_m$) and a field enhancement occurs in the equatorial plane perpendicular to the electric field of light. Such field enhancement will result in an asymmetric ablation around the nanoplasma, leading to formation of an elliptical nanoplasma.

Next, we simulate the intensity distribution of light near an elliptical nanoplasma of $R_x$ = 20 nm; $R_y$ = 40 nm; $R_z$ = 40 nm as illustrated in Fig. 5(b). We observe that significant field enhancement has occurred at the tips of the elliptical nanoplasma (Fig. 5(b)), which promote the growth of the nanoplasma along the the main axis of the elliptical nanoplasma. Our further simulation shows that the field enhancement can also be observed for nanoplasma with larger ellipticities. This is evidenced by Fig. 5(c) which presents the optical intensity distribution in the vicinity of an elliptical nanoplasma of $R_x$ = 20 nm; $R_y$ = 250 nm; $R_z$ = 250 nm. These results indicate that the field enhancement at the tips of nanoplasma is responsible for the continuous growth of SWC with increasing number of the laser pulses, which agrees well with our experimental observation in Fig. 2(e). The growth of single SWC only ceases when it reaches the region where the laser peak intensity in the tight focal spot drops to a level below the ablation threshold.

At the higher laser intensities far above the ablation threshold, the single nanovoid as shown in Fig. 2 can no longer be formed. However, regardless of the focusing geometries, LMZs will be formed in glass due to melting and rapid resolidification. Surrounding the LMZs, there will always be the boundaries which become the interface between two kinds of materials of different optical properties. In some materials, the formed interfaces can provide suitable conditions for surface plasma excitation, which leads to the formation of initial periodic nanovoid chain near the interface.

To model the plasma waves produced at the interface, the optical property of the porous glass modified by the irradiation laser field should be considered. Indeed, the generation of high-density free electron at the interface will lead to a significant change in the dielectric constant of the glass near the interface, which can be calculated using Eq. (1). In addition, similar to the excitation of surface plasma wave with femtosecond laser pulses [22,28], plasma waves can also be excited at the interface, as illustrated in Fig. 6(a). Applying the boundary conditions to the field at the interface would result in the expression for the plasma wave vector as below [29],

$$k_{SPs} = k_0 \sqrt{\frac{\varepsilon_m \varepsilon}{\varepsilon_m + \varepsilon}}, \qquad (2)$$

provided that Re($\varepsilon$) < -$\varepsilon_m$. The electric field at the interface would be periodically enhanced with a spatial period of

$$d = \frac{\pi}{\text{Re}(k_{SPs})}.$$ (3)

Figure 6(b) shows the period $d$ calculated as a function of $N_e$ within the region of Re($\varepsilon$) < $-\varepsilon_m$, which indicates that the minimum plasma density required for excitation of plasma wave is $N_e > 1.5 \times 10^{21}$ cm$^{-3}$. The calculated period is in the range of $d$ = 97~275 nm, which is in good agreement with the nanograting period observed in the experiment. However, we would like to point out that the surface plasma waves excited at the interface do not necessarily directly induce the nanoscale ablation. Instead, the surface plasma waves can induce a periodic distribution of the defects formed near the interface, which facilitates subsequent ablation at the spots with the high-density of defects. The periodically distributed nanovoids are evidenced by the experimental observation as shown in Fig. 4(e). Afterwards, each individual nanovoid can further develop into a SWC, leading eventually to the formation of the PNCA (Fig. 4(i)).

In conclusion, we have revealed the evolution of nanograting formation in a porous glass with the increasing number of irradiation pulses. We stress that nanograting formation is a multi-shot process, which involves the accumulative modification in the volume of glass by irradiation of femtosecond laser and in turn the gradual shaping of the incident laser pulses when they are propagating in the evolving LMZ. Therefore, understanding how the morphology evolves in the volume of glass under the multi-shot irradiation is vital. We observe that surface plasma waves excited at the interface can trigger the formation of nanogratings, which is further supported with our theoretical analyses. Our finding explains why the volume nanogratings formed in glasses share many similar features with the femtosecond-laser-induced surface nanoripples, which also sheds light on the mechanism behind nanostructuring in transparent materials other than the porous glass.

Y. Liao and J. Ni contributed equally in this work. The work is supported by National Basic Research Program of China (No. 2014CB921300) and NSFC (Nos. 61275205, 11104245, 61108015, 61008011, 11174305, 11104294, and 61205209).

**Captions of figures:**

Fig. 1 (Color online) Schematic illustrations of femtosecond laser direct writing setups (a) without and (b) with slit-beam shaping. Insets: close-up views of the nanogratings induced in the written tracks. The laser incident direction (k), polarization direction (E), and the writing direction (S) are indicated in the figure.

Fig. 2 (Color online) Evolution from a single nanovoid to a nanocrack. Scan velocities in (a-f) are 25, 10, 5, 2.5, 1 mm/s, respectively. Scale bar: 1 μm. Position of the focal point is indicated by the dashed lines in each panel.

Fig. 3 (Color online) Cross sectional morphologies of nanogratings written without slit-beam shaping. Scan velocities in (a-f) are 30, 25, 15, 5, 1, 0.95 mm/s, respectively. Scale bar: 1 μm.

Fig. 4 (Color online) Cross sectional morphologies of nanogratings written with slit-beam shaping. Scan velocities in (a-i) are 26, 8, 4.5, 1.25, 0.50, 0.35, 0.10, 0.05, 0.01 mm/s, respectively.

Fig. 5 (Color online) Distribution of intensity of light in the XZ plane near (a) a spherically-shaped nanoplasma (R=20 nm), and two elliptically-shaped nanoplasmas of different sizes (b,c). The sizes of elliptical nanoplasmas are $R_x$=20 nm, $R_y$=40 nm, and $R_z$= 40 nm in (b), and $R_x$=20 nm, $R_y$=250 nm, and $R_z$=250 nm in (c).

Fig. 6 (Color online) (a) Concept of excitation of a surface plasma wave at an interface in the porous glass; (b) the calculated period of the surface plasma wave as a function of electron density. Excitation of surface plasma waves requires $Re(\varepsilon) < -\varepsilon_m$.

Fig. 1

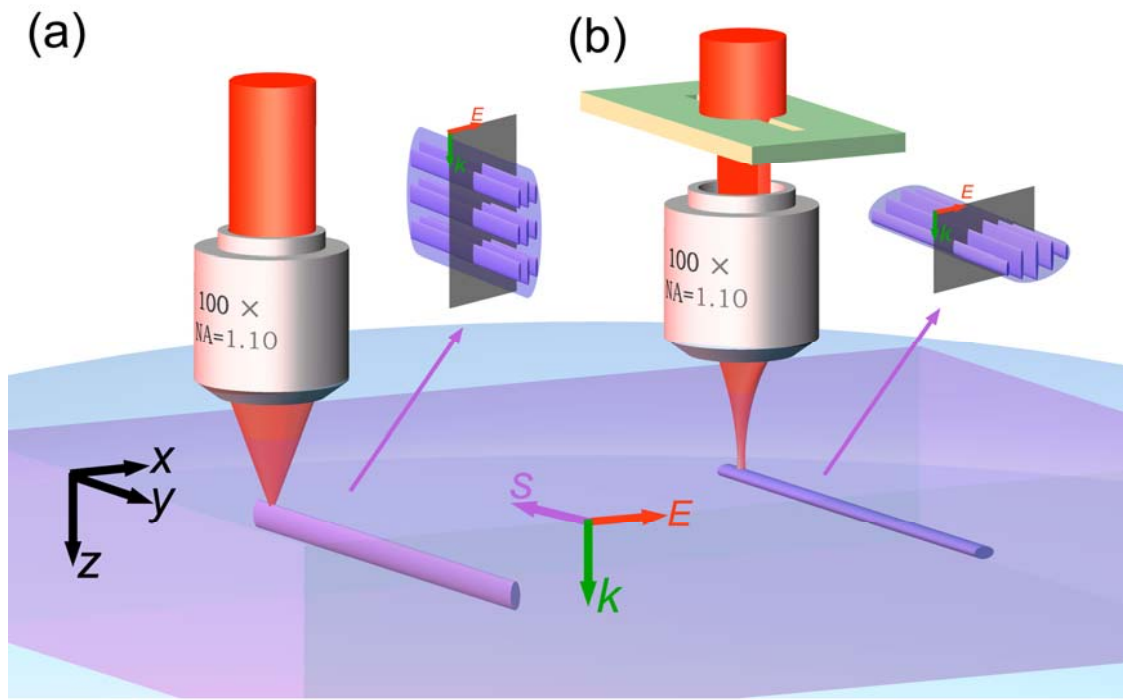

Fig. 2

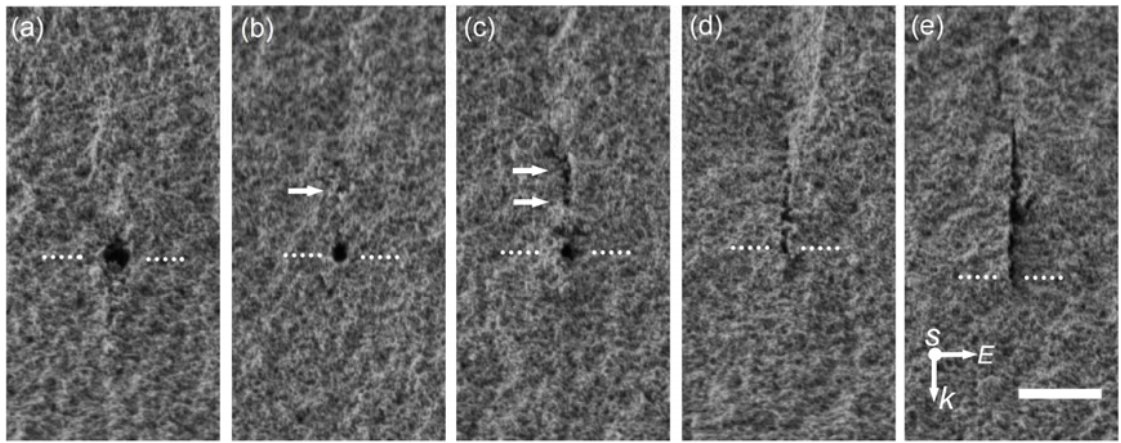

Fig. 3

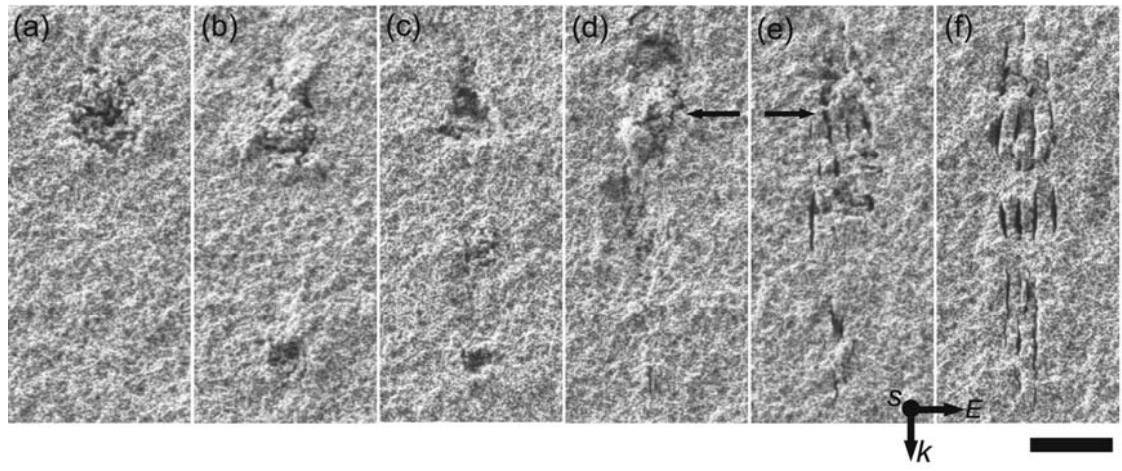

Fig. 4

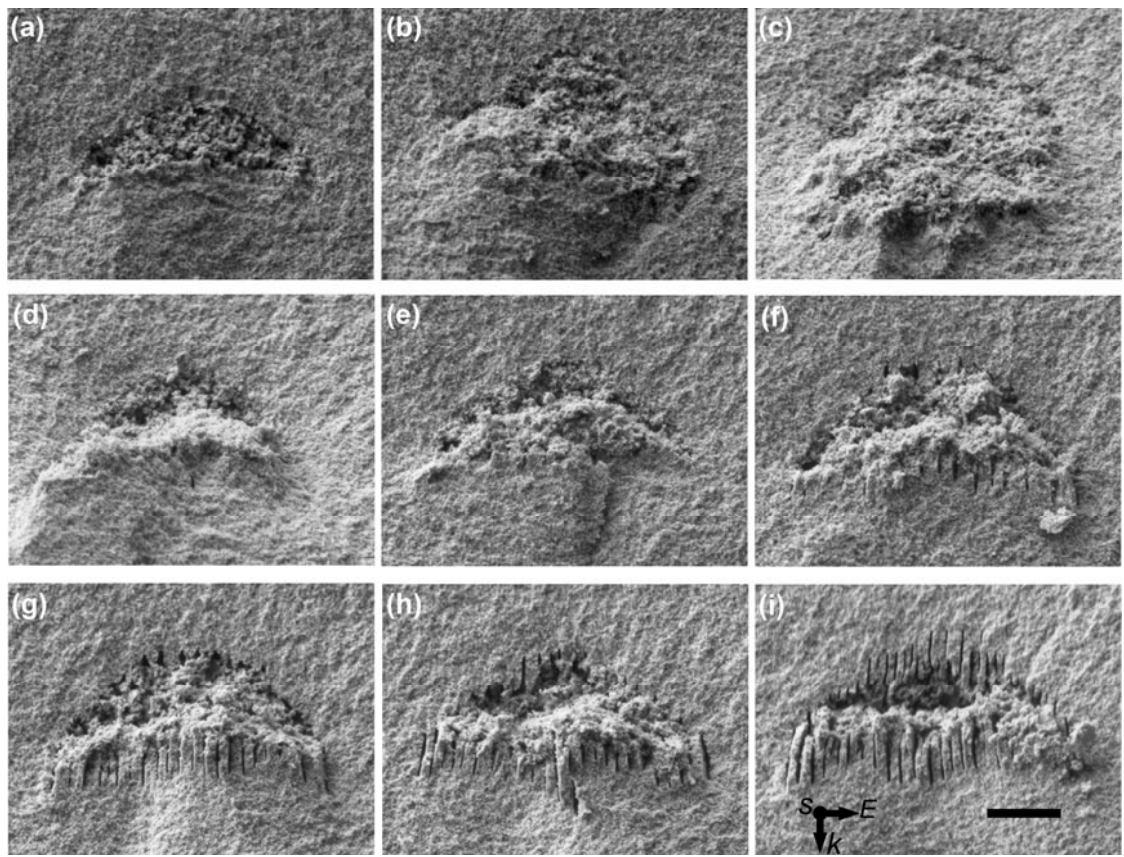

Fig. 5

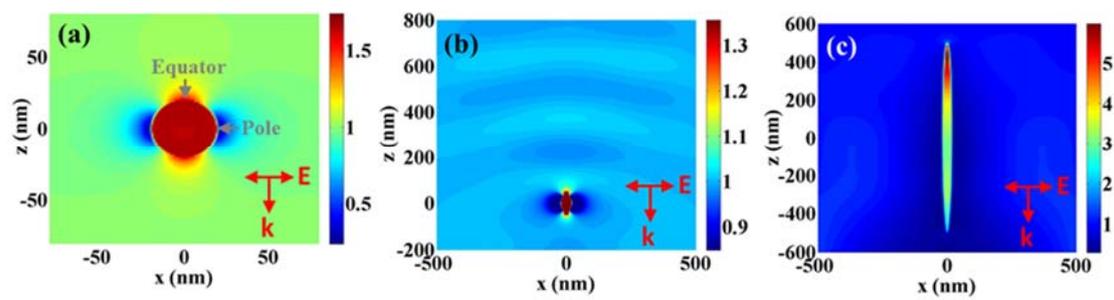

Fig. 6

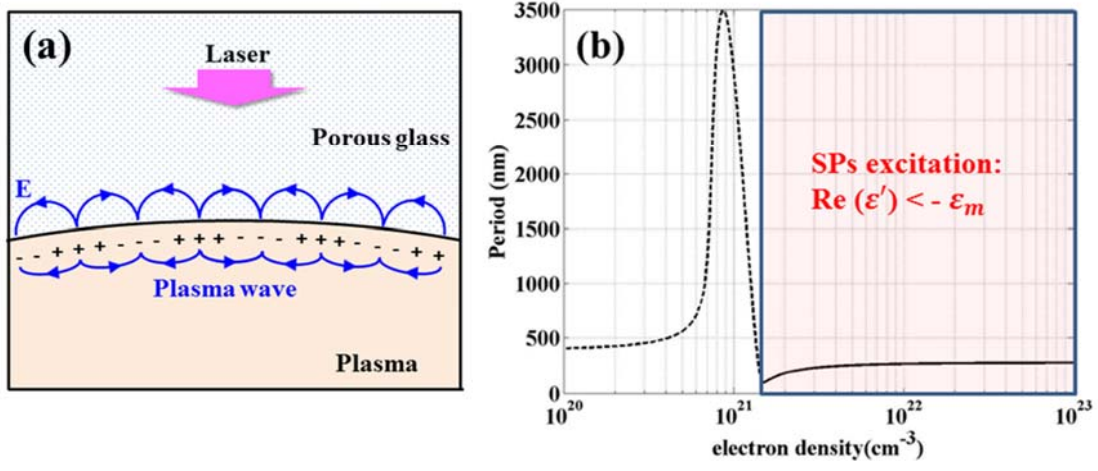